\documentclass{article}

\usepackage{arxiv}

\usepackage[utf8]{inputenc} 
\usepackage[T1]{fontenc}    
\usepackage{hyperref}       
\usepackage{url}            
\usepackage{booktabs}       
\usepackage{amsfonts}       
\usepackage{nicefrac}       
\usepackage{microtype}      
\usepackage{lipsum}
\usepackage{graphicx}
\usepackage{subcaption}
\usepackage[utf8]{inputenc}
\usepackage{amsmath}    
\usepackage{amssymb}    
\usepackage{mathtools}  

\graphicspath{ {./images/} }

\title{A4O: All Trigger for One Sample}
\newcommand{\minisection}[1]{\vspace{0.5\baselineskip}\noindent\textbf{#1.}}
\DeclareUnicodeCharacter{2212}{\ensuremath{-}}

\author{
 Vu Duc Anh, 
 Anh Tuan Tran, 
 Cong Tran, and
 Cuong Pham
}

\begin{document}
\maketitle
\begin{abstract}
Backdoor attacks have become a critical threat to deep neural networks (DNNs), drawing many research interests. However, most of the studied attacks employ a single type of trigger. Consequently, proposed backdoor defenders often rely on the assumption that triggers would appear in a unified way.
In this paper, we show that this naive assumption can create a loophole, allowing more sophisticated backdoor attacks to bypass. We design a novel backdoor attack mechanism that incorporates multiple types of backdoor triggers, focusing on stealthiness and effectiveness. Our journey begins with the intriguing observation that the performance of a backdoor attack in deep learning models, as well as its detectability and removability, are all proportional to the magnitude of the trigger. Based on this correlation, we propose reducing the magnitude of each trigger type and combining them to achieve a strong backdoor relying on the combined trigger while still staying safely under the radar of defenders. Extensive experiments on three standard datasets demonstrate that our method can achieve high attack success rates (ASRs) while consistently bypassing state-of-the-art defenses.
\end{abstract}


\section{Introduction}
Deep neural networks (DNNs) have emerged as the foundational architecture for a wide range of applications, including computer vision \cite{chen2017targeted, gu2017badnets}, natural language processing \cite{dai2019backdoor, kenton2019bert}, as well as fields as diverse as gaming \cite{silver2017mastering, meta2022human} and chemistry \cite{chemistry_dnn, bian2021generative}. The rapid advancements in deep learning have driven significant increases in task complexity, with state-of-the-art models growing progressively deeper and more computationally demanding. Consequently, these models require extensive datasets, which are often challenging to compile, resulting in labeling and training processes that are both intricate and resource-intensive. 
This data dependency frequently compels researchers and practitioners to rely on third-party datasets for model training, delegate training processes to external platforms (e.g., Google Cloud AI Platform, Microsoft Azure Machine Learning), or employ commercial APIs directly for various tasks. 
Such reliance on external resources introduces potential security vulnerabilities, as it creates opportunities for attackers to implant concealed functionalities within DNN models. A body of research demonstrates that DNNs are particularly susceptible to backdoor attacks \cite{gu2017badnets, liu2018trojaning, chen2017targeted}. 
Attackers can introduce a trigger into target models during the training phase by either poisoning a subset of the training data or altering the training procedure. As a result, the compromised model operates as expected on clean inputs but yields specific, adversary-defined outputs when presented with inputs containing the trigger. Such vulnerabilities pose severe risks to critical applications, e.g., face recognition \cite{chen2017targeted}, autonomous driving systems \cite{han2022physical}, and speech recognition \cite{liu2018trojaning}.

Researchers have significantly advanced the field of backdoor attacks by proposing a range of sophisticated techniques, particularly on trigger designs, including patch-based \cite{gu2017badnets}, blending-based \cite{chen2017targeted}, warping-based \cite{nguyen2021wanet}, and signal-based triggers \cite{barni2019new}. Traditionally, most backdoor attack methods have focused on manipulating models using only one type of trigger. However, because each trigger type relies primarily on a single image-processing technique, these approaches are more vulnerable to detection when multiple defense strategies are applied. For instance, Neural Cleanse \cite{neural_cleanse} can expose patch-based and blend triggers, while TeCo \cite{TeCo} identifies triggers by examining the distinctive ways each trigger type interacts with different types of image corruptions. As a result, backdoor attacks that use only one type of trigger are more likely to be detected by contemporary backdoor defense methods. 
Moreover, recent backdoor defenses have leveraged generative models, facilitating more precise and effective isolation and reconstruction of backdoor triggers \cite{BTI-DBF}. These advancements in generative-based defenses have posed increasing challenges for developing resilient backdoor attack methods.

To overcome these limitations, we propose a novel poisoning-based backdoor attack mechanism that aggregates multiple triggers into a united, composite trigger, inspired by the natural phenomenon where combining toxic (or non-toxic) substances can form a more toxic compound \cite{mix_toxic}.
\begin{figure*}[ht]
    \centering
    \includegraphics[width=0.8\linewidth]{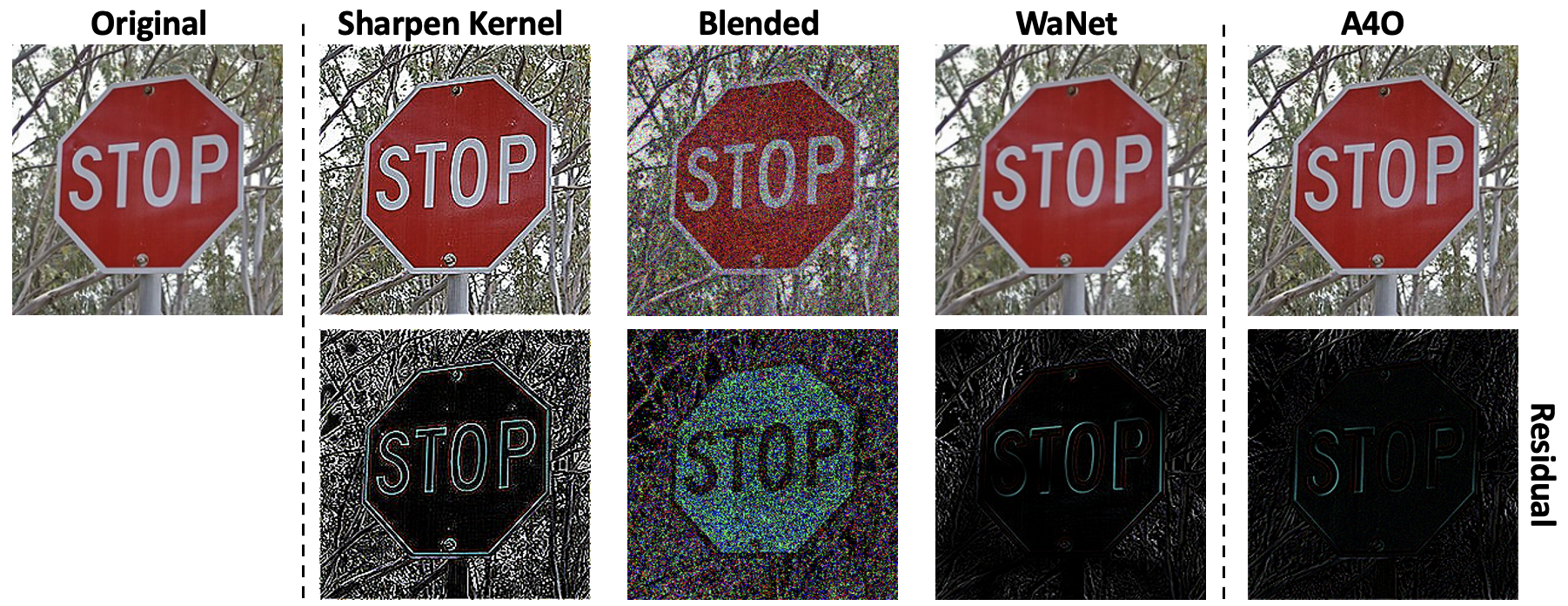}
    \caption{Visualization of backdoor images from different methods. Images on top from left to right: the original image, images generated by Sharpening kernel attack \cite{kernel-attack}, blended backdoor attack \cite{chen2017targeted}, warping-based attack \cite{nguyen2021wanet}, and the proposed A4O attack. Bottom images are residual maps that are amplified by 2$\times$. The images produced by our method appear natural and undetectable, as shown by the residuals.}
    \label{fig:trigger_compare}
\end{figure*}
To obtain a backdoor model, we first formulate the process of finding the trigger components. We identify potential trigger patterns and then proceed to discover the most stealthy variants of these triggers by adjusting their presence in samples. We combine these refined trigger variants to construct a novel and composite trigger. The rationale is that humans are better at recognizing large, consistent abnormalities, but may miss multiple small, inconsistent ones. Since each trigger component is more stealthy than the original, and they are different from each other, our trigger pattern is exceptionally challenging to detect. Note that while there are several prior works \cite{NtoOne, 4frequent, multi-channel-attack, li2024multi} that follow the multi-trigger approach, they simply aggregate different sing-type triggers to boost attack success rate without considering stealthiness. Therefore, these attacks are vulnerable to backdoor defenses, similar or even more fragile than single-trigger attacks. In contrast, our attack is designed with the stealthiness objective in mind; each trigger component is carefully selected and optimized. We provide exemplary backdoor images in Figure \ref{fig:trigger_compare} to showcase the effectiveness of our approach.

Additionally, we introduce an effective method for optimizing these composite triggers through two distinct training modes: \textit{joint mode} and \textit{noise mode}. The joint mode enables the model to learn from all trigger components simultaneously, while noise mode restricts the model’s focus to predefined triggers only. Together, these training schemes yield backdoor models that are both highly effective and difficult to detect. We refer to our method as \textbf{A}ll trigger \textbf{F}or \textbf{O}ne sample Backdoor Attacks (\textbf{A4O-Attack}). Our method is evaluated under two configurations: one with only predefined triggers and another that combines trigger generators with predefined triggers.
In summary, our main contributions are three-fold: 
\begin{itemize}
    \item We propose a novel backdoor attack mechanism that aggregates multiple triggers into one by adjusting the magnitudes of the trigger component to achieve the best balance in terms of effectiveness and stealthiness.
    \item Based on our empirical study, we propose two training modes that ensure backdoor only activated when having all trigger components combined.
    \item Finally, we evaluate our method in two distinct settings: (1) utilizing only predefined triggers and (2) combining predefined triggers with those generated by generative models. In both configurations, our approach achieves state-of-the-art trade-off in attack efficacy and stealthiness, successfully evading existing defense mechanisms.
\end{itemize}

\section{Related Works}
\label{sec:headings}
\subsection{Backdoor Attacks}
Gu et al. \cite{gu2017badnets} introduced the first backdoor attack against deep neural networks (DNNs) by injecting a fixed pixel patch as a trigger into a portion of the training dataset, enabling the model to activate the backdoor upon encountering this trigger. Since the advent of BadNets, numerous backdoor attack methodologies have been developed, often distinguished by the types of triggers used. Examples include the localized patch-based trigger in BadNets \cite{gu2017badnets}, the universal noise-based Blend-triggers \cite{chen2017targeted}, kernel-based \cite{kernel-attack}, warping-based \cite{nguyen2021wanet}, and color-based triggers \cite{color-backdoor}. Other approaches proposed a trigger generator, which utilizes the ability of a generative model to create an adaptive trigger \cite{doan2022marksman, Lira, IAD}. COMBAT \cite{combat} moves forward by utilizing an alternated training to find an appropriate trigger generator for any class. More recent approaches increase attack complexity by integrating multi-trigger and multi-target mechanisms. Some studies incorporate multiple triggers of the same type \cite{NtoOne, 4frequent, multi-channel-attack} by embedding different triggers of the same category into separate image regions. Other studies applied multi-trigger strategies for multi-target attacks \cite{MtoN, NtoOne}, associating distinct triggers with specific target classes, allowing for attacks on any class in the dataset based on the assigned trigger. However, these methods generally retain a single trigger type per class. Recently, Li et al. \cite{li2024multi} proposed a multi-trigger approach that combines multiple triggers within a single sample, primarily to enhance attack success rates without directly addressing the vulnerabilities inherent in multi-trigger configurations.

\subsection{Backdoor Defense}
Alongside the introduction of backdoor attacks, various defenses have been proposed to counter that threat. We can roughly group them into the categories below.

\minisection{Trigger reconstruction based methods.} This approach first tries to approximate the trigger. Then it eliminates the backdoor by suppressing the effect of the reconstructed trigger. The typical defense is Neural cleanse \cite{neural_cleanse}, where a potential trigger is optimized for each class. The model is identified as a backdoor model if there is a class that has a significantly smaller trigger pattern than other classes. Currently, the most advanced trigger reconstruction-based methods were developed in the feature space BTI-DBF \cite{BTI-DBF}, which reconstructs the backdoor trigger by decoupling benign features instead of decoupling backdoor features directly using a generator.

\minisection{Model reconstruction based methods.} These defenses attempt to remove the backdoor by reconstructing or fine-pruning/fine-tuning the infected model. For example, Fine-Pruning \cite{fine_pruning} prunes the potential backdoored neurons based on their average activation values, followed up by revised pruning-based techniques \cite{NAD, channel_prune, shapley_prune}. Meanwhile, ANP \cite{ANP} prunes those neurons that are more adversarial sensitive to remove backdoors. I-BAU \cite{i-bau} proposes to cleanse backdoored models by adversarial training. Recently, RNP \cite{RNP} prunes the backdoored neurons by first unlearning the neurons on the clean defense data and then recovering the clean neurons on the same clean data through filters.

\minisection{Test-time detection methods.} This approach aims to determine whether an inference sample contains a backdoor trigger. For example, STRIP \cite{strip} overlays a series of clean images onto the target image individually and inputs these composites into the model. If the predictions for these overlaid images consistently yield low entropy, the model is flagged as backdoored. Another method, TeCo \cite{TeCo}, assumes that a backdoor-infected model will show distinct robustness to trigger samples under various corruptions, whereas clean images generally maintain similar robustness across corruptions. Recently, MSPC \cite{MSPC} proposed a threshold-free approach that detects backdoors based on the consistency of predictions for trigger samples even when they are scaled, without requiring any additional clean data.

\section{Backdoor Trigger's Magnitude}
\label{sec:btm}
Before introducing our multi-trigger backdoor attack, we first delineate the important findings that we discovered from an assumption: \textit{given a backdoor trigger and a model, the efficiency of a backdoor-infected model will decrease as the magnitude of the trigger decreases gradually}. Here, the trigger's magnitude represents the value of hyperparameters associated with the trigger; higher values correspond to a greater deviation between poisoned and benign samples. For instance, in Wanet's trigger \cite{nguyen2021wanet}, the magnitude is defined by the control-grid size and warping strength, whereas for Blend-trigger \cite{chen2017targeted}, it is defined by the blend ratio. Notably, this assumption applies broadly across various types of backdoors.

\begin{figure*}[!ht]
    \centering
    \includegraphics[width=0.9\linewidth, height = 4cm]{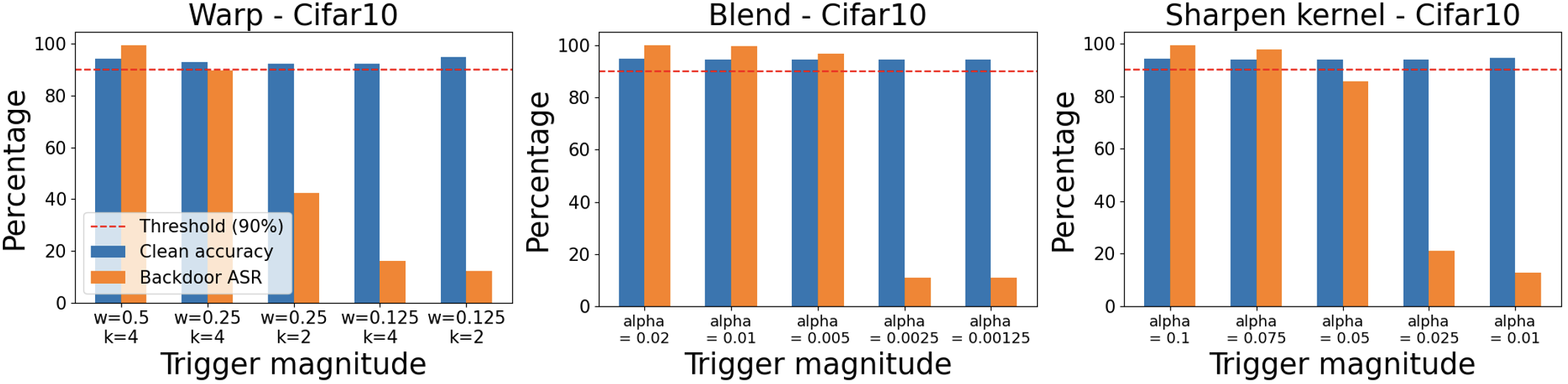}
    \caption{The backdoor-infected model's \textit{clean accuracy} (CA) and \textit{attack success rate} (ASR) when the trigger is reduced with different magnitudes. The decrease indicates that the backdoor-infected models have a consistent correlation between the trigger's magnitude and attack success rate.}
    \label{fig:btm_test}
\end{figure*}

\subsection{Backdoor Trigger's Magnitude Test}
To clearly justify our aforementioned hypothesis, we modify the Corruption Robustness Consistency Test \cite{TeCo} to perform a newly designed test, termed Backdoor Trigger's Magnitude (BTM), which is applicable to various trigger types. Rather than assessing different corruption types at varying levels of severity, this test examines the effects of diverse trigger magnitudes. 

Given a trigger function $\mathcal{B}_K$ with the default trigger magnitude \textit{K} and a backdoor model ${C_{\theta, K}}$ trained with that trigger function, the BTM test computes the clean accuracy (CA) of the clean images and the attack success rate (ASR) of poisoned images with different magnitudes by dividing it by a factor $n \in \mathbb{N}$. The BTM test builds a list ${L_{n}}$  of CA and ASR, where each setup in this list is calculated by: 
\begin{equation}
\textit{L}_{n} =
\begin{cases}
   \frac{1}{S}\sum\limits_{i}^{S}\mathbb{I}({C_{\theta, \frac{K}{n}}({x_i}) = {y_i}})\vspace{3mm}
   \\ 
    \frac{1}{S}\sum\limits_{j}^{S}\mathbb{I}({C_{\theta, \frac{K}{n}}(\mathcal{B}_{\frac{K}{n}}({x_j})) = {y_p})}, 
\end{cases}
\label{eq:BTM_test}
\end{equation}

where $S$ is the number of samples, ${x_i}$ represents a clean sample, $\mathcal{B}_{\frac{K}{n}}$ represents the trigger function with an adjusted trigger magnitude ${\frac{K}{n}}$; ${y_i}$ is the ground-truth label of ${x_i}$, and ${y_p}$ is the target label that the adversaries want the infected model to predict when the trigger sample is given; $\mathbb{I}(.)$ is an indicator function, i.e., $\mathbb{I}(A) = 1$ if and only if \textit{A} is true.

\subsection{Consistency of Trigger's Magnitude}
The list ${L_{n}}$ built in the BTM test can be used to measure the correlation between the magnitude of the trigger, backdoor-infected models' performance, and its visibility against backdoor detection. We choose the default setup of each trigger in their papers as the starting values. We reduce the magnitude by $n$ until the model drops the ASR below 90\%. We conduct the BTM test on three different backdoor attack types: Wanet, Blending, and Kernel with Pre-act Resnet \cite{preact}. From the visualization in Fig.\ref{fig:btm_test}, the ASR of different trigger types degrade due to the degradation of the magnitude of triggers and severely drops when the magnitude is too small. However, many recent defenses \cite{TeCo, MSPC} proved that backdoor triggers are extremely sensitive to image transformation like scaling or corruption. Consequently, the trigger should not be too small in order to maintain its stability.

\section{Proposed Method}
\vspace{-0.1cm}
We first introduce our threat model and then introduce the proposed backdoor attacks.
\vspace{-0.2cm}
\subsection{Threat Model}
In this study, we adopt a data poisoning and threat model based on previous works \cite{chen2017targeted, nguyen2021wanet, hidden_backdoor}, where an attacker supplies a poisoned dataset to a victim for training. The attacker has two primary objectives. The first, centered on effectiveness, is to ensure that the backdoored model achieves a high attack success rate (ASR) by consistently classifying poisoned inputs into the target class label(s), while maintaining high performance on clean, non-poisoned inputs. Additionally, this effectiveness goal requires that the model generalizes well in handling poisoned inputs across diverse datasets. The second objective is stealthiness, aiming for the backdoored model and poisoned dataset to evade detection and removal by human analysts or backdoor defenses. In practical scenarios, however, victims may employ various backdoor defenses that can detect and mitigate different types of backdoors. As a result, existing backdoor attacks may often fail in real-world settings.

To address this limitation, we propose a more robust backdoor attack mechanism by aggregating multiple types of backdoor triggers. By adjusting the strength of each trigger, this approach enhances resilience against a wide range of defenses.
\vspace{-0.1cm}
\subsection{All-For-One Backdoor Attack}
Our All-For-One backdoor attack can be considered a complex, extended type of single-trigger backdoor attack since we still follow one trigger for all datasets. However, unlike the presented backdoor trigger, which is established by a solid pattern, our trigger combines multiple triggers into one. Here, we demonstrate the method overview, trigger selection, and poisoning procedure to build backdoor datasets.
\vspace{-0.4cm}
\subsubsection{Overview}
\begin{figure}[t]
    \centering
    \includegraphics[width=0.7\linewidth, height = 5.3cm]{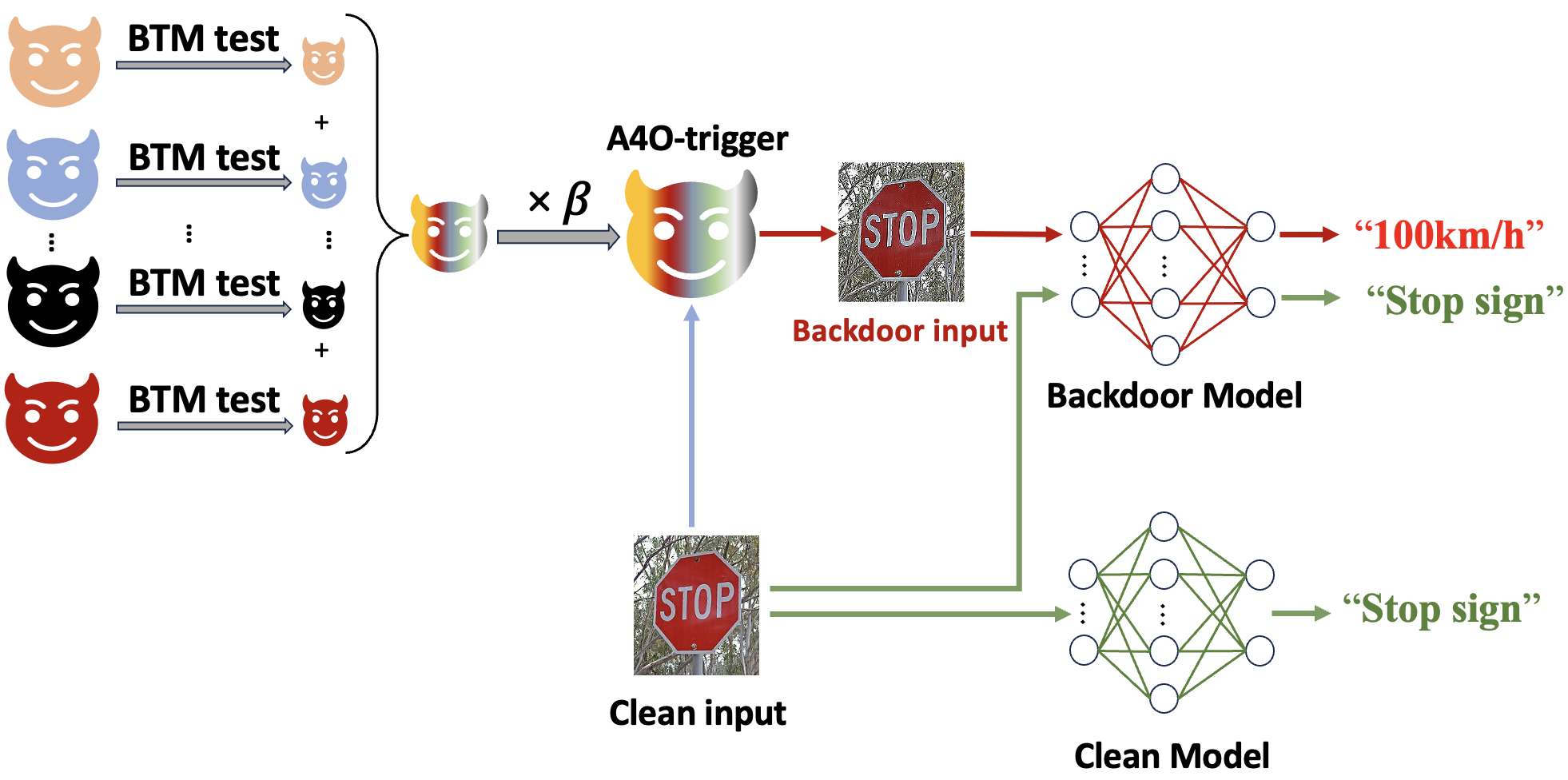}
    \caption{An illustrative of A4O backdoor attack. We combine multiple triggers with adjusted magnitudes to get an efficient and stealthy composited attack.}
    \vspace{-0.3cm}
    \label{fig:mbda_general}
\end{figure}

We focus on image classification tasks. Given an image classification dataset $\mathcal{D} = {\{(x_i, y_i)|i = 1,...,N\}}$. Attackers generate backdoor samples with a trigger function $\mathcal{B}: \mathcal{X} \to \mathcal{\overline{X}}$ where $\mathcal{X}$ and $\mathcal{\overline{X}}$ are clean and poisoned data samples, respectively. For each clean pair $({x_i}, {y_i})$, trigger function will map the sample $x_i$ into a backdoor sample ${x_p}$, i.e, ${x_p} = \mathcal{B}(x_i)$ and modifies its label to target label ${y_p}$. To ensure the stealthiness and efficacy of backdoor injection, the adversary randomly chooses a few training samples to poison, which creates a set of backdoor samples ${\mathcal{D}_p} = {\{(x_p, y_p)\}}$. The poisoned dataset becomes $\hat{\mathcal{D}} = {{\mathcal{D}_c} \cup {\mathcal{D}_p}}$, where ${\mathcal{D}_c} = ({x_c},{y_c})$ represents 
 remaining clean data, yielding a poisoning rate $\alpha = |{\mathcal{D}_p}|/|{\hat{\mathcal{D}}}| $. However, differing from previous backdoor attack works, our main focus is to design a multi-trigger function that is based on multi-type triggers. The multi-trigger function can be formulated as:
\begin{equation}
    {x_p} = {\mathcal{B}_m}(...({\mathcal{B}_2}({\mathcal{B}_1}(x_i)))), 
\label{eq:MTBA_eq}
\end{equation}
where ${\mathcal{B}_i}$ is a backdoor trigger function and $m$ is the number of different backdoor trigger functions applied to clean sample ${x_i}$. An illustration of the proposed multi-trigger backdoor, along with the details
of the backdoor attack, is described in Figure \ref{fig:mbda_general}.

\subsubsection{Trigger Selection}
As previously discussed, to evade current backdoor defense mechanisms, our main idea is to design a heterogeneous trigger by combining multiple triggers with distinct properties into a unified form. To construct this complex trigger, we carefully select various types of triggers commonly used in mainstream backdoor attacks, such as patch-based triggers like BadNet \cite{gu2017badnets}, geometric distortion triggers like WaNet \cite{nguyen2021wanet}, and blending triggers like Blend-trigger \cite{chen2017targeted}, among others. The selected triggers are required to meet the following two criteria: \begin{itemize} 
\item The chosen backdoor triggers must differ significantly from one another, exhibiting distinct characteristics. 
\item The triggers must allow for adjustable intensity, enabling control over their prominence within the sample. 
\end{itemize}
\vspace{-0.1cm}
\subsubsection{Poisoning Procedure}
As in Section \ref{sec:btm}, triggers with lower magnitudes are less detectable and pose a reduced risk to the model. Conversely, triggers can enhance the model's ASR by increasing their magnitudes, although this may lead to easier detection. In such cases, it is plausible to consider that triggers can boost their magnitudes and ASR by aggregating various smaller triggers without raising detectability concerns. 
To implement this strategy, firstly, we determine the trigger's lowest magnitude to achieve over 90\% ASR, and to prevent it from image transform manipulation, we scale it to a $\beta$ scale. The magnitude scale formula can be interpreted as:
\begin{equation}
\begin{aligned}
    \hat{\mathcal{B}}_i = \mathcal{B}_i * \beta
\end{aligned}
\label{eq:Trigger_magnitude}
\end{equation}
We then randomly sample a few training samples into a backdoor candidate subset ${\mathcal{D}_p}$. For each sample in ${\mathcal{D}_p}$, we apply Equation \ref{eq:MTBA_eq} to sequentially poison the samples with each trigger from the selected triggers. 
\vspace{-0.1cm}
\subsubsection{Training Modes}
Since the A4O-attack incorporates multiple types of triggers, it may be vulnerable to detection by patch-based or noise-adding defenses. To mitigate this, we adopt an approach similar to that used in Wanet \cite{nguyen2021wanet}, applying randomized transformations (e.g., warping, noise, kernel alterations) to each selected trigger. These transformations differ from the predefined triggers applied to clean samples, allowing the network to bypass backdoor activation and correctly predict the intended class. We term this as $\textit{noise mode}$, which is defined as follows:
\begin{equation}
\begin{aligned}
(x_i,y_i) \mapsto (\mathcal{B}_i(x_i, \text{uniform}_{[-1, 1]}), y_i), \quad i =[1..m] \\
\end{aligned}
\label{eq:noise_mode}
\end{equation}
in which $\text{uniform}_{[-1, 1]}(.)$ is a function returning a random tensor with the trigger shape and element value in the range $[−1, 1]$.

Another issue that we observe is biased learning, where the backdoored model may predominantly rely on a single backdoor trigger component, effectively reducing it to a single-trigger backdoor attack model. To address this, we propose a novel training approach termed \textit{joint mode}. In joint mode, we poison clean samples using only one of the component triggers, while retaining the original sample label. This process prevents the model from activating the backdoor when only a single component trigger is present. Consequently, the model is encouraged to learn the backdoor only from the aggregated trigger, promoting an even distribution of influence across all component triggers. The \textit{joint mode} can be formulated as:
\begin{equation}
\begin{aligned}
(x_i,y_i) \mapsto (\mathcal{B}_i(x_i), y_i), \quad i =[1..m]  \\
\end{aligned}
\label{eq:joint_mode}
\end{equation}
We first select the backdoor probability $p_b \in (0,1)$ and the noise probability $p_n \in (0,1)$ as well as joint probability $p_j \in (0,1)$ with $p_b + p_n + p_j < 1$. 
Accordingly, training a dataset with A4O attack can be formulated as follows:
\begin{equation}
\scalebox{0.9}{$
(x_i, y_i) \mapsto 
\begin{cases} 
(x_i, y_i) & \text{with probability } 1 - p_b - p_n - p_j \\
(\mathcal{B}(x_i), y_p)) & \text{with probability } p_b \\
(\mathcal{B}_i(x_i, \text{uniform}_{[-1, 1]}), y_i) & \text{with probability } p_n \\
(\mathcal{B}_i(x_i), y_i) & \text{with probability } p_j,
\end{cases}
$}
\label{eq:Training_mode}
\end{equation}
where the $\mathcal{B}$ is aggregated trigger. We noticed that these training modes are efficient through various defenses. The details are discussed in the Ablation studies in Section \ref{Ab_stu}.

\section{Experimental Results}
\label{result}
\vspace{-0.1cm}
\subsection{Implementation Settings}
\begin{figure}[h!]
    \centering
    \begin{subfigure}{0.5\textwidth}
        \includegraphics[width=\linewidth]{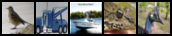}
        \caption{Original images}
    \end{subfigure}
    
    \vspace{0.2em}

    \begin{subfigure}{0.5\textwidth}
        \includegraphics[width=\linewidth]{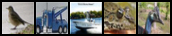}
        \caption{Multi-trigger poisoned images}
    \end{subfigure}
    \caption{Visual comparisons of the original images and multi-triggered images from CIFAR10.}
    \vspace{-0.3cm}
    \label{fig:result_image}
\end{figure}

\vspace{-0.2cm}
\minisection{Datasets} We conduct backdoor attack experiments on four benchmark datasets, including Cifar10\cite{cifar10}, CelebA\cite{celeba}, and (a subset of ImageNet) Tiny-ImageNet\cite{tiny_image} with 200 classes. Note that the CelebA dataset has annotations for 40 independent binary
attributes, which is not suitable for multi-class classification. Therefore, we follow the configuration suggested by \cite{salem2022dynamic} to select the top three most balanced attributes, including Heavy Makeup, Mouth Slightly Open, and Smiling, and then concatenate them to create eight classification classes. 

\minisection{Network training} We used ResNet18\cite{preact} as the victim model backbone. We trained the networks using the SGD optimizer. The initial learning rate was 0.01, which was reduced by a factor of 10 after each 100 training epochs. The networks were trained until convergence. We use $\beta = 5/3$, $ p_b = 0.1$, $p_n = 0.05$ and $p_n = 0.05$ for each trigger. For the trigger component, we use $k = 0.25$ and $s = 4$ for Wanet, blend ratio $\alpha = 0.005$ for noise-blended, and blend ratio $\gamma = 0.075$ for Sharpen Kernel.
\subsection{Attack Experiments}
\begin{table*}[h]
    \centering
    \resizebox{0.9\textwidth}{!}{%
    \begin{tabular}{ccccccccccccc}
        \toprule
         Attack $\rightarrow$    & \multicolumn{2}{c}{A4O} & \multicolumn{2}{c}{A4O-Generator} & \multicolumn{2}{c}{Combat} & \multicolumn{2}{c}{Wanet*} & \multicolumn{2}{c}{Blend*} & \multicolumn{2}{c}{Kernel*}\\
        \cmidrule(lr){2-3} \cmidrule(lr){4-5} \cmidrule(lr){6-7} \cmidrule(lr){8-9} \cmidrule(lr){10-11} \cmidrule(lr){12-13} 
        Dataset $\downarrow$& {ACC} & {ASR}  & {ACC} & {ASR}  & {ACC} & {ASR} & {ACC} & {ASR} & {ACC} & {ASR} & {ACC} & {ASR} \\ 
        \midrule
        CIFAR10      & 94.72 & 99.96 & 94.23 & 99.98 & 94.49 & 100 & 93.03 & 90.72 & 94.94 & 10.62 & 94.14 & 93.77\\
        CelebA       & 79.46 & 99.77 & 79.31 & 99.97 & 79.43 & 99.87 & 79.21 & 93.50 & 79.44 & 99.59 & 79.42 & 98.99\\
        Tiny-Image   & 58.91 & 100   & 59.88 & 99.99 & 59.41 & 100     & 59.64 & 99.96 & 48.95 & 10.87 & 59.81 & 99.94\\
        \bottomrule
    \end{tabular}
    }
    \caption{\textbf{Attack experiments.} The * indicates that this method is the sub-trigger from our multiple triggers. Wanet* is the Wanet trigger with $k=4$,$s=0.25$. Blend* is the Noise blend trigger with blend ratio $\alpha=0.005$. Kernel* is the Sharpen filter trigger with a blend ratio $\gamma=0.075$.}.
    \vspace{-0.3cm}
    \label{tab:attack_experiment}
\end{table*}

We train and test the backdoor models in an all-to-one configuration, in which the target label ${y_p}$ is the same for all images. We follow two setups. In the first setup, we combine three pre-defined triggers: WaNet, Noise blended trigger, and Sharpening kernel trigger. In the second setup, we combine two pre-defined triggers, noise-blended trigger, and  Sharpening kernel trigger, with a generator-based version of WaNet, following the COMBAT mechanism \cite{combat}. We apply $\textit{joint mode}$ on all datasets and $\textit{noise mode}$ on CelebA, Tiny-Image due to their complexity. The clean accuracy (ACC) and attack success rate (ASR) are shown in Table \ref{tab:attack_experiment}. As can be seen, with clean images, the networks could correctly classify them like any benign models, with an accuracy of around 94\% on CIFAR-10, 59\% on Tiny-ImageNet, and 79\% on CelebA. When applying the multi-trigger, the attack success rate was near 100\% on all three datasets (CIFAR10, Tiny-ImageNet, and CelebA). However, when using a single trigger from the sub-triggers, the classifiers trained with A4O attacks only have the ASRs around 20-40\%, as shown in the Supplementary, confirming that the models only activate backdoor when all sub-triggers appear together. In Table \ref{tab:attack_experiment}, we also include the results of COMBAT \cite{combat}, a state-of-the-art baseline, and single-trigger attack experiments. A4O variants have equivalent performance compared to COMBAT while consistently outperforming the single-trigger ones. Since our method degrades each component trigger before combining, the poisoned images look almost identical to the original, effectively bypassing human inspection, as shown in Figure \ref{fig:result_image}.
\vspace{-0.1cm}
\subsection{Defense Experiments}
In this section, we evaluate the backdoor-injected model against popular and recent defense mechanisms. 

\begin{figure*}[h]
    \centering
    \begin{subfigure}{0.75\textwidth}
        \centering
        \includegraphics[height=3cm, width=12cm]{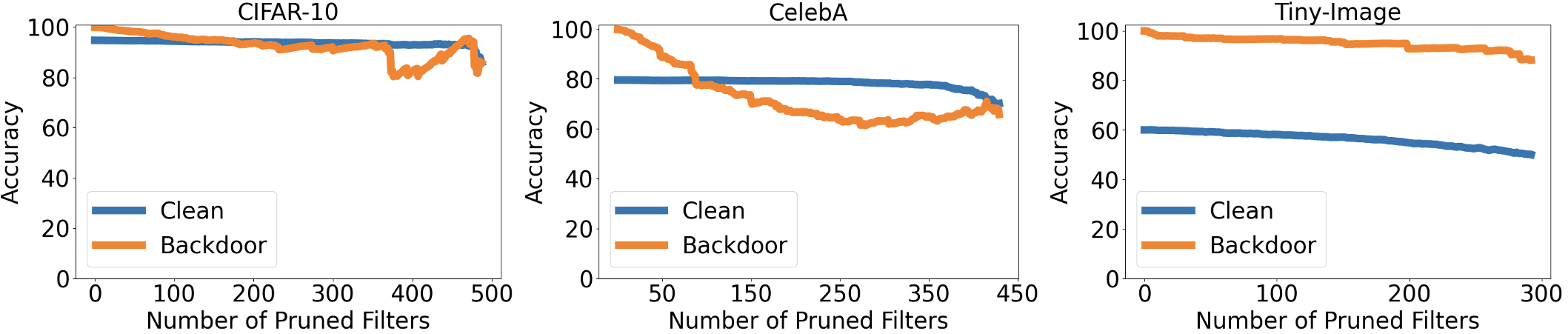}
        \caption*{(a) A4O-predefined trigger}
        \label{fig:pruning_predefine}
    \end{subfigure}
    \begin{subfigure}{0.24\textwidth}
        \centering
        \includegraphics[height=2.9cm]{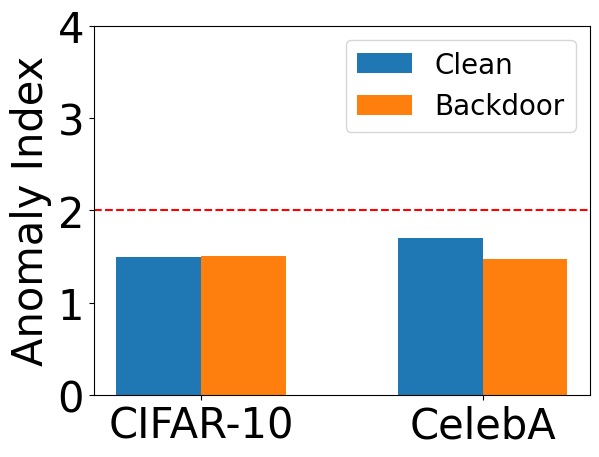}
        \caption*{(c) A4O-predefined trigger}
        \label{fig:neural_cleanse_predefine}
    \end{subfigure}
    \begin{subfigure}{0.75\textwidth}
        \centering
        \includegraphics[height=3cm, width=12cm]{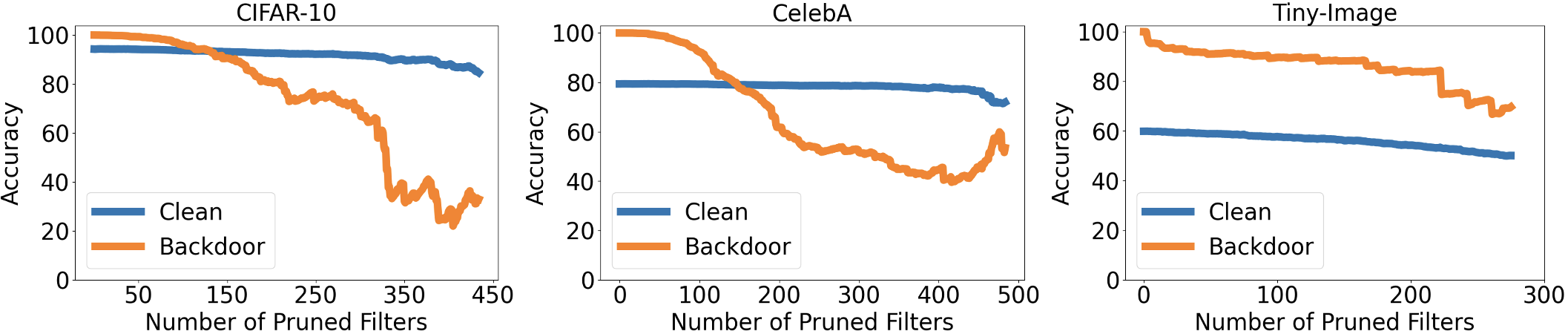}
        \caption*{(b) A4O-generator trigger}
        \label{fig:pruning_generator}
    \end{subfigure}
    \begin{subfigure}{0.24\textwidth}
        \centering
        \includegraphics[height=2.9cm]{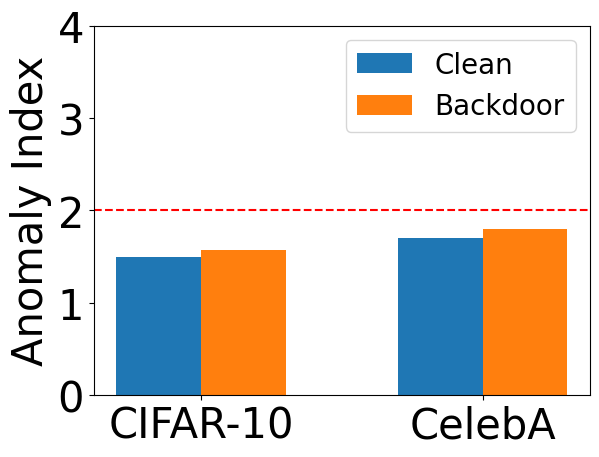}
        \caption*{(d) A4O-generator trigger}
        \label{fig:neural_cleanse_generator}
    \end{subfigure}
    \captionsetup{justification=centering}
    \caption{Models' performance against Fine-pruning (a, b) and Neural cleanse (c, d).}
    \label{fig:fine-pruning}
\end{figure*}

\minisection{Neural Cleanse (NC)} \cite{neural_cleanse} is a model-defense method based on the pattern optimization approach. It assumes that the backdoor is patch-based. For each class, NC computes the optimal patch pattern to convert any clean input to that target label. It then checks if any label has a significantly smaller pattern as a backdoor sign. NC quantifies it by the Anomaly Index metric with the clean/backdoor threshold $\tau$ = 2. Since our trigger is based on multiple types (warp, blend, kernel), a different mechanism compared with the patch pattern created by NC, our multi-trigger passed the test on CIFAR-10 and CelebA (Figure  \ref{fig:fine-pruning}). We notice that the Tiny-ImageNet dataset is already considered a naturally triggered dataset \cite{natural_trigger} and we also tried NC on the clean model which shows an anomaly index over 2. Therefore, we do not include Tiny-ImageNet in this defense.

\minisection{Fine Pruning (FP)} \cite{fine_pruning} and \textbf{RNP} \cite{RNP}, instead, focus on neuron analyses. For FP, given a specific layer, it analyzes the neuron responses on a set of clean images and detects the dormant neurons, assuming they are more likely to tie to the backdoor. These neurons are then gradually pruned to mitigate the backdoor. We tested FP on our models and plotted the network accuracy, either clean or attack, with respect to the number of neurons pruned in Figure \ref{fig:fine-pruning}. The pruning stops when the ACC drops more than 10\%. We notice that pruning cannot completely remove our backdoor. For RNP, it first unlearns the clean neurons by maximizing the model’s error on a small subset of clean samples and then recovers the neurons by minimizing the model’s error on the same data by learning a pruning mask on the model. The potential backdoor neural will be filtered to mitigate the backdoor. Prior multi-trigger backdoor attack \cite{li2024multi}, as showed in their experiments, was completely  exposed by this defense. In contrast, A4O-attacks completely surpass RNP and also outperforms other methods in this test (Table \ref{tab:RNP_defense}). 

\begin{table*}[h]
    \centering
    \resizebox{0.9\textwidth}{!}{%
    \begin{tabular}{ccccccccccccc}
        \toprule
         Attack $\rightarrow$    & \multicolumn{2}{c}{A4O} & \multicolumn{2}{c}{A4O-Generator} & \multicolumn{2}{c}{Combat} & \multicolumn{2}{c}{Wanet*} & \multicolumn{2}{c}{Blend*} & \multicolumn{2}{c}{Kernel*}\\
        \cmidrule(lr){2-3} \cmidrule(lr){4-5} \cmidrule(lr){6-7} \cmidrule(lr){8-9} \cmidrule(lr){10-11} \cmidrule(lr){12-13} 
        Dataset $\downarrow$& {ACC} & {ASR}  & {ACC} & {ASR}  & {ACC} & {ASR} & {ACC} & {ASR} & {ACC} & {ASR} & {ACC} & {ASR} \\ 
        \midrule
        CIFAR10      & 36.62 & 00.00  & \textbf{12.59} & \textbf{64.06} & \underline{36.22} & \underline{9.05} & 93.27 & 19.04 & - & - & 82.71 & 5.94\\
        CelebA       & \textbf{27.16} & \textbf{100}   & 27.16 & 100 & 27.16 & 100 & 27.16 & 100   & 27.16 & 100  & 27.16 & 100\\
        Tiny-Image   & 6.14  & 65.91   & \underline{9.89} & \underline{90.98} & \textbf{0.57} & \textbf{99.97} & 51.22 & 6.62  & - & - & 36.78 & 0.05\\
        \bottomrule

    \end{tabular}
    }
    \caption{A4O backdoor attack against RNP defense. The best results are in \textbf{bold}. The second best results are in \underline{underline}. We highlight that our method perfectly surpasses RNP on CelebA, Tiny-ImageNet, and CIFAR10.} 
    \label{tab:RNP_defense}
\end{table*}

\begin{figure*}[ht]
    \centering
    \begin{minipage}{0.6\textwidth}
        \centering
        \includegraphics[width=\linewidth]{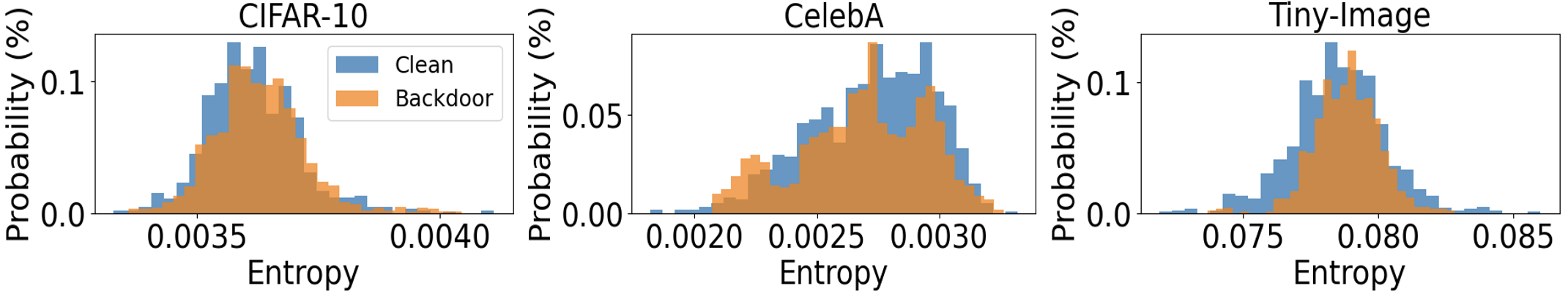}
        \vspace{1pt}
        \includegraphics[width=\linewidth]{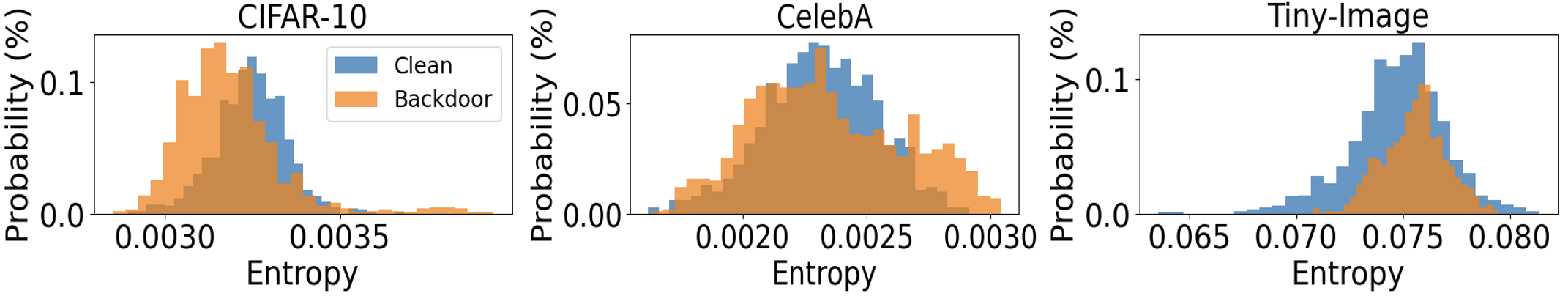}
        \caption{A4O (top) and A4O-Generator (bottom) against STRIP.}
        \vspace{-0.2cm}
        \label{tab:STRIP}
    \end{minipage}
    \hfill
    \begin{minipage}{0.35\textwidth}
        \centering
        \small
        \begin{tabular}{ccc}
            \toprule
            Attack $\rightarrow$  & A4O & A4O-Generator \\
            \cmidrule(lr){2-3}
            Dataset $\downarrow$ & AUROC$\downarrow$ & AUROC$\downarrow$ \\ 
            \midrule
            CIFAR10      & 0.68 & 0.74 \\
            CelebA       & 0.6 & 0.63 \\
            Tiny-Image   & 0.48 & 0.56 \\
            \bottomrule
        \end{tabular}
        \captionof{table}{A4O backdoor attack against MSPC defense. This defense is ineffective against our method.}
        \vspace{-0.2cm}
        \label{tab:MSPC_defense}
    \end{minipage}
\end{figure*}

\minisection{STRIP} \cite{strip} can detect a backdoored model if the predictions of superimposed input images exhibit persistence with low entropy. Figure \ref{tab:STRIP} shows the results of STRIP. A4O-attack expresses similarities between entropy distributions of clean and backdoored samples. 

\minisection{TeCo} \cite{TeCo} and \textbf{MSPC} \cite{MSPC}, instead, focus on image transformation. TeCo suggests that triggered samples will behave differently when subjected to various corruptions, whereas benign samples will behave consistently. This highlights the need for a strong, resilient trigger to overcome this defense. Conversely, MSPC indicates that triggered samples will perform consistently across different scale-ups, unlike clean samples. This shows that a robust trigger cannot bypass MSPC. Together, these two defenses form an effective combination for detecting backdoor samples. The result is shown in Table \ref{tab:TeCo_defense} and Table \ref{tab:MSPC_defense}. Our trigger, which is composed of multiple triggers, can minimize the possibility of being affected by various image errors, and each component trigger is not strong enough to counter the scaling mechanism. 

\begin{table}[ht]
    \centering
    \resizebox{0.48\textwidth}{!}{%
    \begin{tabular}{ccccc}
        \toprule
        Attack $\rightarrow$   & \multicolumn{2}{c}{A4O} & \multicolumn{2}{c}{A4O-Generator} \\
        \cmidrule(lr){2-3} \cmidrule(lr){4-5}    
        Dataset $\downarrow$  & {AUROC$\downarrow$} & {F1 Score$\downarrow$} & {AUROC$\downarrow$} & {F1 Score$\downarrow$} \\ 
        \midrule
        CIFAR10      & 0.16 & 0.5  & 0.46 & 0.56 \\
        CelebA       & 0.46 & 0.55 & 0.35 & 0.58 \\
        Tiny-Image   & 0.61 & 0.63  & 0.27 & 0.5 \\
        \bottomrule

    \end{tabular}}
    \caption{A4O backdoor attack against TeCo defense. Our method completely surpass this defense.}
    \vspace{-0.3cm}
    \label{tab:TeCo_defense}
\end{table}

\minisection{BTI-DBF} \cite{BTI-DBF} differs from other defenses by focusing on benign features rather than directly targeting backdoor features, eliminating the need for trigger knowledge. It first isolates benign features from backdoor ones using only benign samples. Then, it trains a trigger generator to minimize differences in benign features while maximizing them in poisoned features. This generator is subsequently used in backdoor removal and pre-processing defenses. This defense demonstrates significant superiority when confronting pre-defined trigger attacks. This is because the triggers generated by pre-defined trigger attacks often deviate substantially from the benign images. However, the BTI-DBF method shows limitations when confronted with sequential training approaches that utilize a trigger generator, such as COMBAT. The COMBAT technique effectively optimizes the triggers' features to closely mimic clean features, leading to BTI-DBF's inability to distinguish between these features. The results against BTI-DBF are shown in Table \ref{tab:BTI-DBF_defense}.

\begin{table*}[h]
    \centering
    \begin{subtable}{0.4\textwidth}
        \centering
        \tiny 
        \renewcommand{\arraystretch}{0.9} 
        \resizebox{\textwidth}{!}{%
        \begin{tabular}{ccccc}
            \toprule
             Attack $\rightarrow$    & \multicolumn{2}{c}{A4O} & \multicolumn{2}{c}{A4O-Generator} \\
            \cmidrule(lr){2-3} \cmidrule(lr){4-5}  
            Dataset $\downarrow$& {ACC} & {ASR}  & {ACC} & {ASR}  \\ 
            \midrule
            CIFAR10      & 79.35 & 1.79  & 81.09  & 6.41 \\
            CelebA       & 76.67 & 8.98  & 76.71  & 48.38\\
            Tiny-Image & 16.13 & 0.17 & 19.48 & 0.27  \\
            \bottomrule
        \end{tabular}
        }
        \caption{A4O backdoor attack against BTI-DBF(U) defense.}
        \label{tab:BTI-DBF_defense1}
    \end{subtable}
    \hspace{0.3cm}
    \begin{subtable}{0.4\textwidth}
        \centering
        \tiny 
        \renewcommand{\arraystretch}{0.9} 
        \resizebox{\textwidth}{!}{%
        \begin{tabular}{ccccccccccccc}
            \toprule
             Attack $\rightarrow$    & \multicolumn{2}{c}{A4O} & \multicolumn{2}{c}{A4O-Generator} \\
            \cmidrule(lr){2-3} \cmidrule(lr){4-5} 
            Dataset $\downarrow$& {ACC} & {ASR}  & {ACC} & {ASR} \\ 
            \midrule
            CIFAR10    & 90.89 & 0.79  & 91.4   & 67.52 \\
            CelebA     & 68.50 & 10.12 & 70.84  & 22.08 \\
            Tiny-Image & 56.59 & 4.03  & 56.83  & 71.72 \\
            \bottomrule
        \end{tabular}
        }
        \caption{A4O backdoor attack against BTI-DBF(P) defense.}
        \label{tab:BTI-DBF_defense2}
    \end{subtable}
    \caption{A4O backdoor attack against BTI-DBF defenses. The results show that our method can surpass this defense.}
    \vspace{-0.3cm}
    \label{tab:BTI-DBF_defense}
\end{table*}

\vspace{-0.3cm}
\subsection{Abalation Studies}
\label{Ab_stu}
\vspace{-5pt}
\minisection{Number of triggers} We extend our study by using a two-trigger components setup to strengthen our approach. In our experiments, we employ a combination of two component triggers—warping and blending—referred to as Two4One. This method maintains high performance in terms of Clean Accuracy and Attack Success Rate while effectively evading various defense mechanisms. These results demonstrate the stealthiness and robustness of our proposed methods. The Two4One's results against defenses can be found in the Appendix.

\minisection{Magnitude Scale} We investigated the effect of the magnitude scale $\beta$. We compare the proposed method with two settings, the original magnitude of each component trigger (we call it Multi-trigger) and the A4O with $\beta = 1$ (we call it A4O-based). We conducted a test on CIFAR10 and Tiny-Image against state-of-the-art model reconstruction-based defense (RNP) to indicate our superiority. The result is shown in Table \ref{tab:RNP_baseline}.

\begin{table}[ht]
    \centering
    \resizebox{0.49\textwidth}{!}{%
    \begin{tabular}{ccccccc}
        \toprule
         Attack $\rightarrow$ & \multicolumn{2}{c}{A4O} & \multicolumn{2}{c}{A4O-based} & \multicolumn{2}{c}{Multi-trigger} \\
        \cmidrule(lr){2-3} \cmidrule(lr){4-5} \cmidrule(lr){6-7}
        Dataset$\downarrow$ & {ACC} & {ASR}  & {ACC} & {ASR} & {ACC} & {ASR}  \\
        \midrule
        CIFAR10     & 36.62 & 00.00  & 93.92 & 0.98  & 93.69 & 0.72 \\
        Tiny-image  & 6.14 & 65.91  & 5.47 & 91.42  & 20.61 & 100 \\
        \bottomrule
    \end{tabular}
    }
    \caption{Compare between A4O and base-line multi-trigger backdoor attack. Our method possesses consistent results through different datasets compared with baseline.}
    \vspace{-0.3cm}
    \label{tab:RNP_baseline}
\end{table}

\minisection{Role of noise mode and join mode} Without the noise mode, we could still train a backdoor model with similar clean and attack accuracy. However, these models have lower performance on TeCo, as shown in Table \ref{tab:TeCo_defense_noise}. Optimized trigger patterns and additional noises revealed their true behavior.
\vspace{-0.2cm}
\begin{table}[h]
    \centering
    \resizebox{0.49\textwidth}{!}{%
    \begin{tabular}{ccccccc}
        \toprule
        Attack $\rightarrow$    & \multicolumn{2}{c}{w 2 mode} & \multicolumn{2}{c}{w/o joint mode} & \multicolumn{2}{c}{w/o noise mode} \\
        \cmidrule(lr){2-3} \cmidrule(lr){4-5} \cmidrule(lr){6-7}                 
        Dataset$\downarrow$ & {AUROC$\downarrow$} & {F1 Score$\downarrow$} & {AUROC$\downarrow$} & {F1 Score$\downarrow$} & {AUROC$\downarrow$} & {F1 Score$\downarrow$} \\ 
        \midrule
        CIFAR10      & 0.26 & 0.5  & 0.21 & 0.51 & 0.16 & 0.5  \\
        CelebA       & 0.46 & 0.55 & 0.66 & 0.63 & 0.69 & 0.65 \\
        Tiny-Image   & 0.61 & 0.63  & 0.52 & 0.58 & 0.99 & 0.96 \\
        \bottomrule

    \end{tabular}}
    \caption{A4O backdoor attack against TeCo defense with and without join/noise mode. The results show the efficiency of our proposed training modes.}
    \label{tab:TeCo_defense_noise}
\end{table}
\section{Conclusion}
This paper introduces a novel backdoor attack method that poisons multiple triggers in one sample by optimizing their magnitude. We examine our attack mechanism in two setups: predefined trigger only and generator-based trigger combined with pre-defined triggers. We then incorporate in training two novel modes (noise mode and joint mode) making it stealthy and passing all the known defense methods. Results on benchmark datasets verified the effectiveness of our methods and their performance against different defenses that posed a significant threat to DNNs. Finally, investigating the connection between the trigger elements and the compromised classifier will be valuable for enhancing backdoor defense research.

\bibliographystyle{unsrt}  
\bibliography{references}  

\end{document}